\documentclass[12pt,preprint]{aastex}

\usepackage{emulateapj5}

\slugcomment{To appear in {\it The Astrophysical Journal} }
\shorttitle{CO Luminosity Function and $\Omega_{HI+H_2}$}
\shortauthors{Keres, Yun \& Young}

\begin{document}

\title{CO Luminosity Functions For FIR and B-band Selected Galaxies \\
and the First Estimate for $\Omega_{HI+H_2}$}

\author{Dusan Keres, Min S. Yun \& J. S. Young}
\affil{Department of Astronomy, University of Massachusetts, Amherst, MA
01003}
\email{keres@nova.astro.umass.edu, myun@astro.umass.edu, young@astro.umass.edu}

\begin{abstract}
        
We derive a non-parametric CO luminosity function using a 
FIR and an optical $B$-band selected sample of the galaxies
included in the FCRAO Extragalactic CO Survey.  The FIR
selected sample is defined using the 
IRAS Bright Galaxy Surveys (BGS; IRAS 60 micron flux density $\ge 5.24$ Jy). 
Although our CO sample is not complete, the normalization using the
BGS reproduces the IRAS 60 micron luminosity function in 
excellent agreement with those found in the literature.  Similarly,
a $B$-band selected sample defined using the Revised Shapley-Ames
(RSA) catalog is used to derive a CO luminosity function for a comparison.
A Schechter function describes the both derived CO luminosity 
functions reasonably well.
Adopting the standard CO-to-H$_2$ conversion factor, we derive
a molecular gas density of $\rho_{H_2}=(3.1\pm 1.2) \times 10^7h 
M_\odot$ Mpc$^{-3}$ for the local volume. 
Combining with the measurements of the local HI mass density and the
helium contribution, we estimate that the total mass density of 
cold neutral gas in the local universe is 
$\Omega_{gas} =(4.3 \pm 1.1)\times 10^{-4} h^{-1}$,
which is about 20\% of the total stellar mass density $\Omega_*$.

\end{abstract}


\section{Introduction \label{sec:intro}}

The star formation history of the Universe is closely linked with
the evolution of the gas content of the Universe.  Observations of
the distribution and total gas content in galaxies and in intergalactic
clouds offer some of the most important observational constraints for the
cosmology and galaxy evolution models. Because stars form out of cold,
dense gas clouds, the history of galaxy formation and evolution is also the
history of gas accretion and conversion into stars.
Here we derive the total cold gas density for the local volume
by deriving the local CO luminosity function and combining with
the existing estimates of the neutral atomic gas density. 

A variety of luminosity functions (LFs) and mass functions are found
in the literature: optical \citep[e.g.][]{mar94}, 
HI \citep[e.g.][]{zwa97,sch98}, and infrared \citep [e.g.][]{soi87,yun01}. 
While several groups have investigated the local neutral  
hydrogen density, little information is available on the
molecular gas content in galaxies because of the lack of
appropriate data. In her pioneer work, \citet{ver87}  
derived averaged values of CO luminosity and the CO/HI flux ratio 
based on a maximum likelihood probability distribution 
for 40 galaxies and 47 upper limits obtained by combining all 
available data at the time.  Since the large fraction of the
CO measurements used were non-detections (upper limits), the derivation
of the CO luminosity function was highly problematic.  Nevertheless,
a general trend of increasing number of CO emitting galaxies
per unit volume with decreasing CO luminosity was suggested by this 
analysis. 

Nearly 15 years later, the availability of the extragalactic CO 
data has improved greatly.  The FCRAO (Five College Radio 
Astronomy Observatory) Extragalactic CO Survey
\citep{yng95} represents a particularly rich database for the
investigation of CO luminosity function and molecular
gas content in galaxies.  We have constructed a large, statistically
significant sample of far-infrared (FIR) and optical $B$-band 
selected galaxies from this survey and derived a
non-parametric CO luminosity function.  Our sample galaxies range
over 4 orders of magnitudes in CO luminosity, and only a small
fraction are non-detections.
Since CO is a tracer of hydrogen molecules, molecular hydrogen mass
can be derived from CO luminosity. By integrating the 
resulting molecular gas mass function, we then derive
the molecular gas mass density and the total cold gas mass
density of the local universe.

\bigskip
        
\section{Sample Selection and Properties \label{sec:sample}}

The FCRAO Extragalactic CO Survey (``the Survey'', herein) 
is the largest CO survey of galaxies available in the literature, 
containing observations of 300 galaxies at 1421 positions \citep{yng95}. 
Since the Survey covers a broad range of Hubble types and
contains a large number of detections (236 or 79\%), it is well suited 
for a statistical analysis and for constructing a CO luminosity function.
The majority of the survey galaxies were selected from the Second 
Reference Catalog \citep[RC2;][]{dvv76} or the IRAS database. Most of the
galaxies in the Survey are spiral or irregular galaxies at
declination north of $-25^\circ$ and satisfy at least one of the following
three criteria: $B^0_T < 13$, $S_{60\mu m} > 5$ Jy, or $S_{100\mu m} > 10$ Jy 
\citep[see][for a detailed description]{yng95}. 
The majority of the detected galaxies in the Survey were observed at 
multiple positions along the major axis, and the total CO fluxes 
were derived by modeling the underlying CO distribution. 

The Survey is not complete in terms of any of the sample selection 
criteria, and the full sample cannot be used directly for a statistical
analysis in a straightforward way.  The analysis of the velocity 
integrated CO luminosity is further complicated by the fact that 
the CO line width varies from one galaxy to another (and even one
position to another in a given galaxy) 
while the sensitivity achieved also varies depending
on the observing condition even if identical integration
times are used (typically 2-4 hours).  Therefore, it is 
difficult to address the sensitivity and completeness limit 
of the Survey in terms of the observed CO flux. 
Instead, taking advantage of the well-known tight correlation
between the FIR and CO luminosity \citep[see][]{yng95}, 
we utilize the IRAS 60 $\mu$m flux density of 
the individual sample galaxies to define a complete sample
for a statistical analysis and the derivation of the CO LF.
We further test the robustness of the technique and the
selection bias by deriving the CO luminosity function using the optical
$B$-band selection (see \S~\ref{sec:COBLF}).
Previously \citet{br93} successfully derived the HI mass function 
for the field galaxies using the optical selection, and the 
radio luminosity function has been successfully derived using 
the FIR selection function by \citet{yun01}. 

Taking advantage of the well defined and well studied IRAS Bright
Galaxy Samples \citep[BGSs;][]{soi89,san95}, we adopt the
sample selection criterion of 60 $\mu$m flux density limit greater 
than 5.24 Jy.
A total of 200 galaxies in the Survey satisfy this criteria. 
Most of the selected galaxies are spirals of types Sa-Sc while 14 galaxies are 
mergers, 7 galaxies are close pairs, and 3 galaxies have no type determined.
In most cases we adopt the 60 $\mu$m fluxes from the BGS surveys.
For galaxies with angular diameters larger than 8\arcmin, 
we adopt the values from \citet{ric88} after multiplying by 1.18 
in order to match the flux scaling \citep [see][]{dey90}. This rescaling
has little impact on our sample completeness limit because most
of our sample galaxies have flux densities much larger than 5.24 Jy. 
Some of the 60 $\mu$m flux density measurements come from 
\citet{yng89,yng02}, and 10 galaxies satisfying our selection criterion 
are added to the sample.  Five of these galaxies are 
not in the BGS survey area, and two are present in \citet{ric88} 
but not in the BGS surveys.

There are 12 galaxies in in our selected sample that are brighter than 
5.24 Jy at 60 $\mu$m and were not detected in CO 
(see Table~\ref{tab:missing_galaxies}).  The CO (1--0) flux measurements
for five of these galaxies are found in the literature.
The remaining 7 galaxies are treated as two limiting cases: (a)
as detections with zero flux; and (b)
as detections at the upper limit flux value \citep[see][]{ver87}. 
A large number of non-detections would severely limit the determination
of the the true CO LF, but the non-detections account for less 
than 4\% in our sample and thus have little overall impact.

Some of the galaxy distances come from direct 
measurements such as using Cepheids.  We adopt a Hubble constant of
$H_0=75$ km s$^{-1}$ Mpc$^{-1}$ for the remaining galaxies. 
Since we are using the CO database obtained from largely a single 
instrument (2.5\% of the sample taken from other surveys), the
internal consistency of the data and the analysis should be
quite good.  Our sample also includes 26 galaxies in the Virgo cluster. 
We adopt a uniform distance of 16 Mpc for the Virgo galaxies.
We have constructed a CO LF with and without Virgo
galaxies in order to see if the presence of the cluster galaxies
influenced the shape of the CO LF (see \S~3).  

\begin{figure}[ht]
\plotone{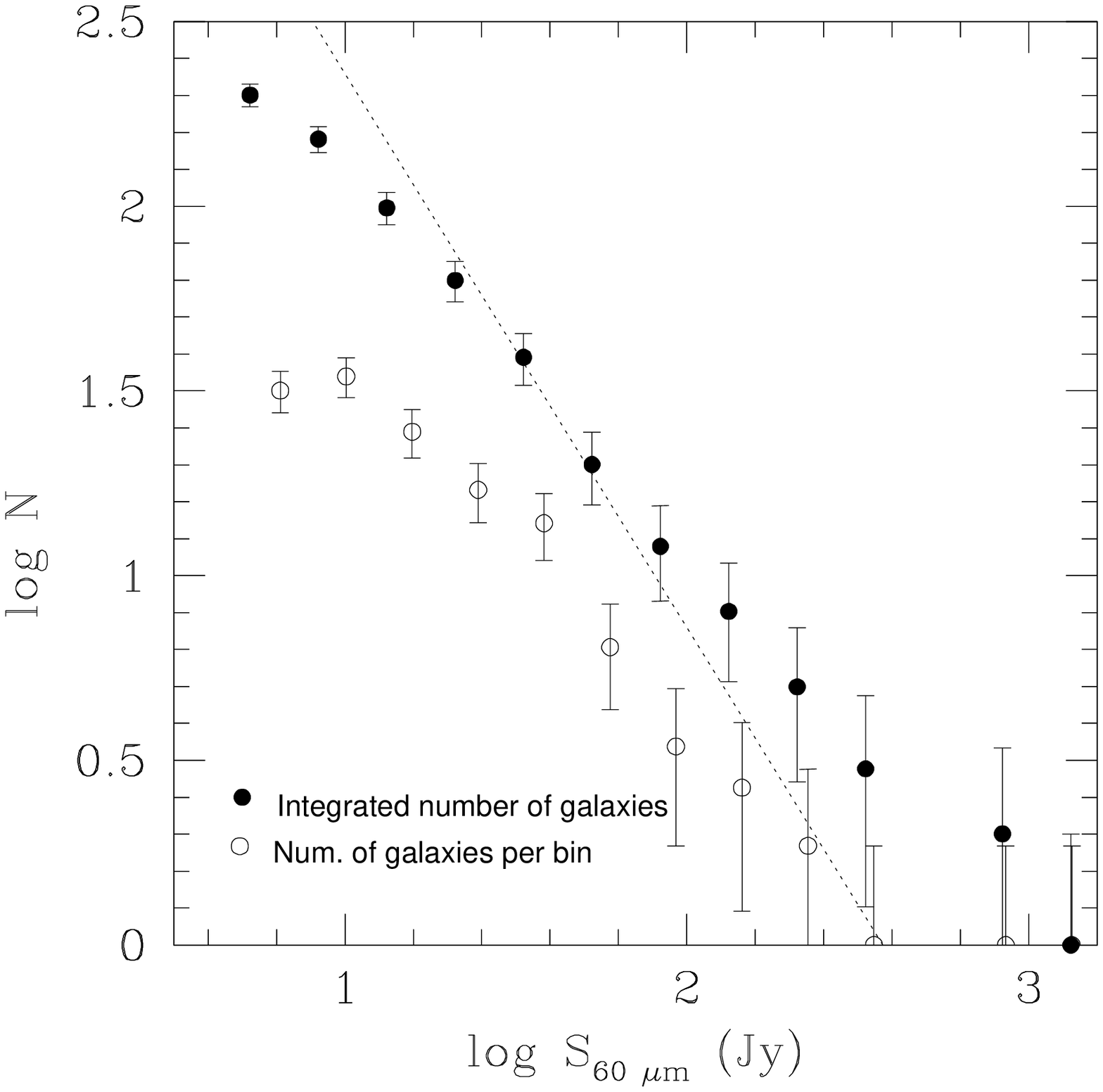}
\caption{
Number of galaxies per bin (0.2 in log scale) of 60 $\mu$m flux
density and integrated number of galaxies. Although the number of galaxies per
bin grows with decreasing flux density, the power-law slope of 
$\alpha \sim -1$ suggests a non-uniform distribution of sources.
The dotted line represents the expected behavior of uniformly distributed
non-evolving sources in Euclidean geometry, appropriate for the
601 galaxies in the two BGS samples. The fact that high flux bins 
are higher than the complete sample line is noticed also in BGS \citep{soi89} 
and is caused by large scale structures. The lowest luminosity bins 
are lower than expected probably because sources near the flux
cutoff that are spatially resolved by IRAS have systematically lower
flux entries in the Point Source Catalog and are thus missed.
\label{fig:S60counts}
}
\end{figure}

To examine the uniformity and completeness of our sample, we have
analyzed the differential number count statistics as a function of
the $60 \mu m$ flux density (see Fig.~\ref{fig:S60counts}). 
If the $60 \mu m$ sources are uniformly distributed in a Euclidean
space and are not evolving, the resulting differential number 
count should be a power-law distribution with $N\propto S^{-3/2}$.  
The number of sample galaxies per bin shown in 
Figure~\ref{fig:S60counts} is consistent with such a power-law,
but the slope is somewhat shallower.  When all of the flux density
bins are included, the best fit power-law index is $\alpha=-1.0$,
where $log~N=N_0+\alpha\times log~S$.  For the $60 \mu m$ 
flux density between 10 and 100 Jy where most of the sample 
galaxies fall, the differential number counts are more consistent
with the uniform distribution within the statistical uncertainties.

\begin{figure}[ht]
\plottwo{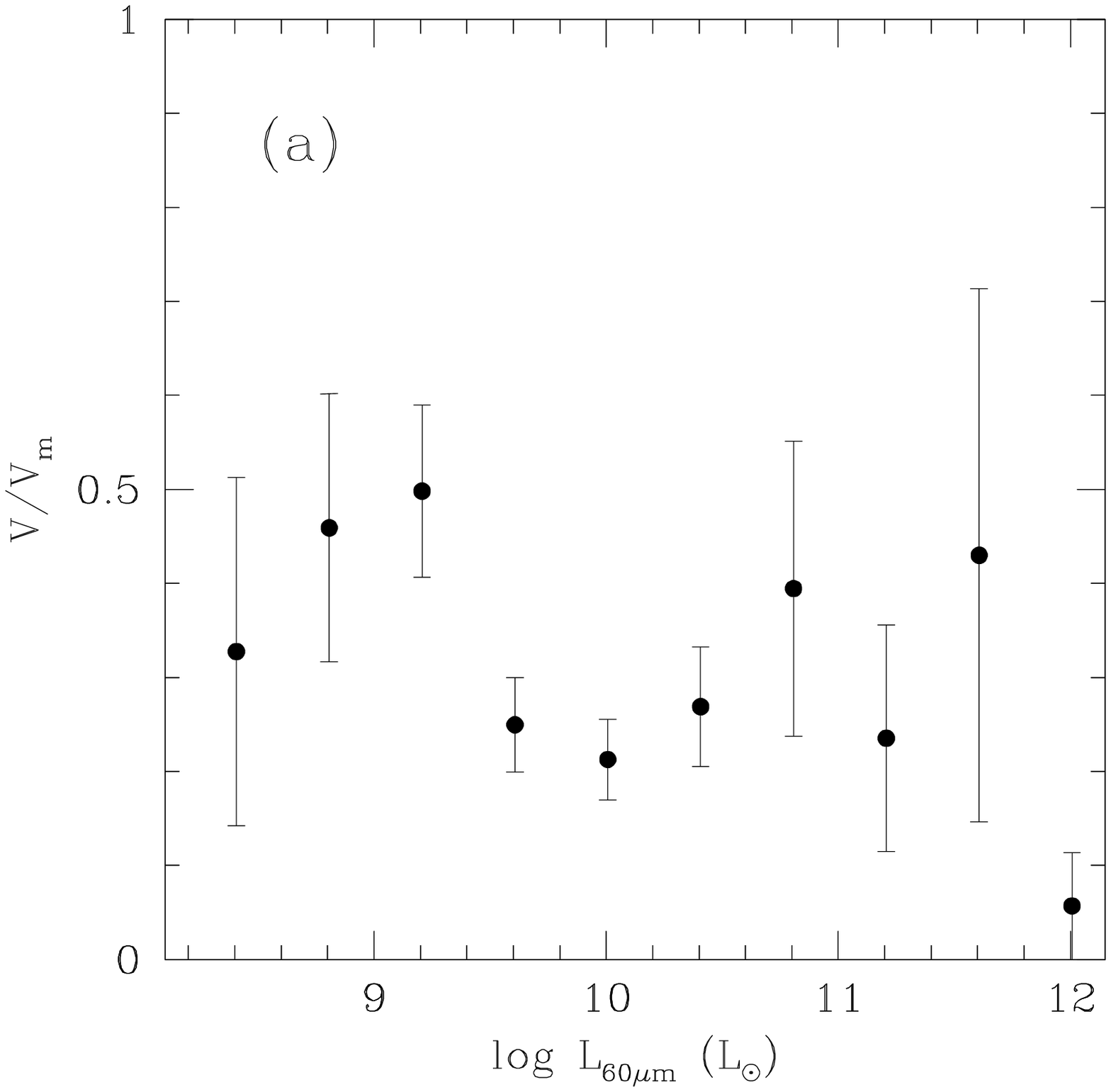}{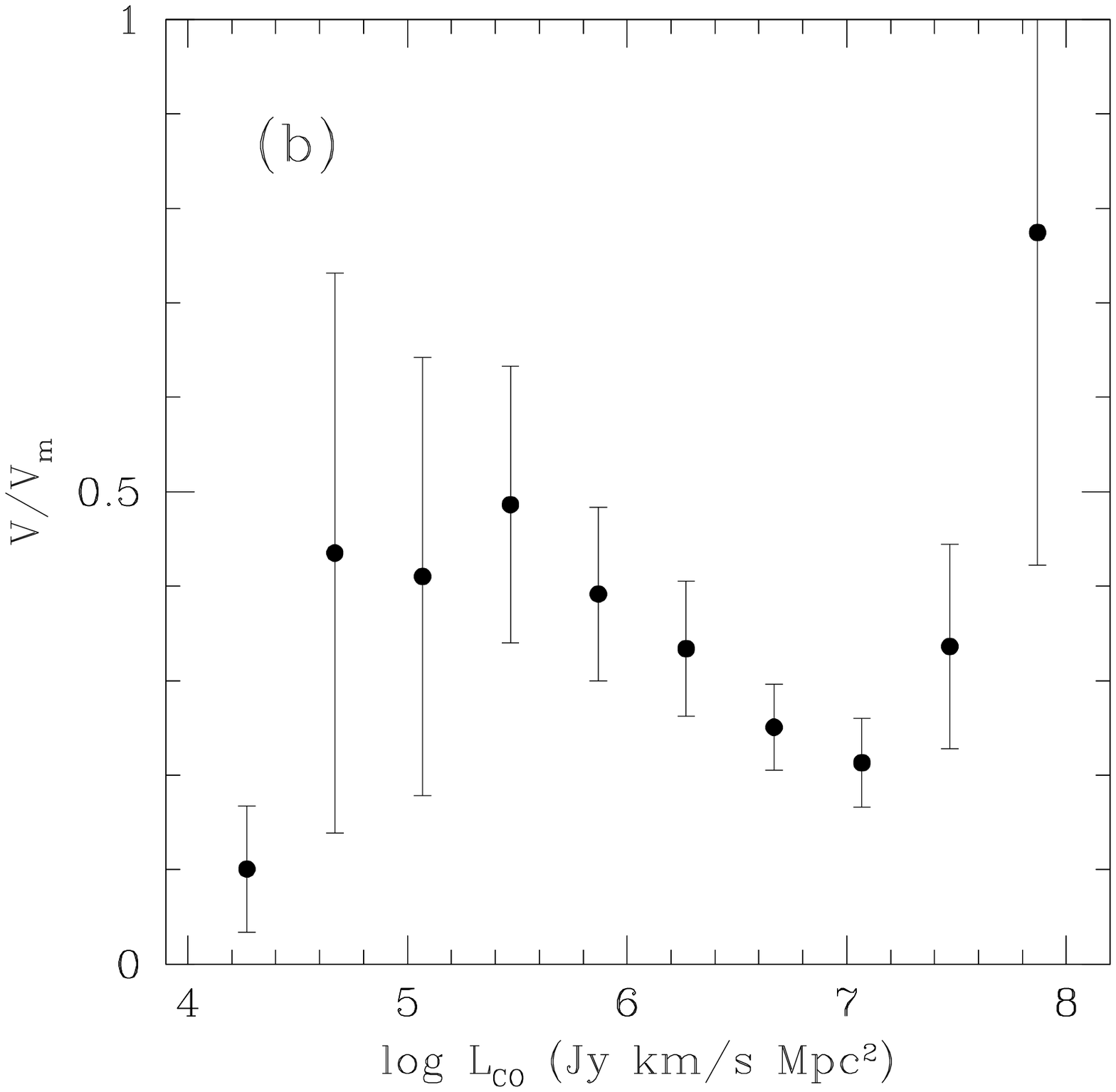}
\caption{
$V/V_m$ analysis for (a) the 60 $\mu$m LF and (b) the CO LF. 
The mean value of $V/V_m\sim 0.35$ suggests that our sample 
galaxies are on average about 10\% closer than the uniform case. 
\label{fig:vvm}
}
\end{figure}

There are several factors contributing to the flatter 
than expected power-law slope in the observed number counts
in Figure~\ref{fig:S60counts}.  First of all, the spatial 
distribution of galaxies is in general {\it not} uniform, particularly
given the relatively high flux density cutoff we adopted.
The situation is exacerbated by the fact that our Galaxy
is located within a galaxy aggregate called the Local 
Supercluster.  The flattening of the galaxy
counts among the high flux density bins as seen in 
Figure~\ref{fig:S60counts} is an immediate outcome of the
local large scale structures.  \citet{soi89} noted a similar
number count enhancement in their analysis of the first 
Bright Galaxy Sample (BGS1).  
The mean $<V/V_m>$ for our sample is about 0.35, 
indicating that on average our sample galaxies are about 10\%
closer than the uniform case.  The $<V/V_m>$ values are particularly
low for the log $L_{60\mu m}$ bins of 9.6-10.4 (see Fig.~\ref{fig:vvm}),
and similar trend is also seen in the $<V/V_m>$ plot by \citet{soi89}.
Comparison to \citet{soi89} value $<V/V_m>=0.47$ is suggesting that major
cause of this discrepancy is weak bias in our sample towards galaxies closer
than average expected from the complete sample, while the large scale 
structure plays less important role. Although the FCRAO CO Survey
selected target galaxies ``at random'' 
from the parent sample of IRAS and $B$-band selected galaxies,
a slight bias favoring galaxies which were well suited to 
the FCRAO resolution (1$'$-5$'$ in size) is also present.  
Such a bias manifests 
non-trivially in these statistics, however.  Given the
limitation of our sample, such as selecting galaxies suitable for
observation with FCRAO telescope and the local large scale structures, 
the steep power-law index of $\alpha\sim -1.2$ shown in 
Figure~\ref{fig:S60counts} suggests that our assumption of
random and uniform sampling from the parent samples still seems reasonable.
Effects of these potential biases on the derivation of the LFs are evaluated
by the derivation of the 60$\mu m$ LF using the same sample, as discussed
bellow.
\bigskip

\section{Luminosity Functions \label{sec:LFs}}

A luminosity function represents the space density of the sample galaxies 
per luminosity bin ($\Delta L$) centered on L. It 
represents the probability density of finding a galaxy 
with a specific luminosity, and it also contains information on 
the total luminosity in the sampled volume. We use the classical $1/V_m$ 
method \citep{sch68} to derive the LF and associated uncertainty 
directly as:
\begin{equation}
\rho (L)=\sum^N_{i=1} {1 \over V_m^i}, \ \ \ \ 
\sigma_\rho(L)=(\sum^N_{i=1} {1 \over (V_m^i)^2})^{1/2}
\label{eq:1/vm}
\end{equation}
\noindent 
where $\rho (L)$ is number of objects per volume per luminosity bin centered 
on luminosity L, $V_m$ is the sample volume appropriate for each of the galaxies, 
and N is the number of galaxies in the bin.

The FCRAO Extragalactic CO Survey covers the same area as the IRAS
BGS \citep{soi89,san95}, and we selected a subset using the same
IRAS 60 $\mu$m flux density cutoff.  Therefore, the sampling 
correction for our sample can be reduced to a simple scaling relation 
if our sample is a fair
subset of the BGS, as argued above (\S~\ref{sec:sample}).
In this case, the scaling factor is the ratio of the number of galaxies
contained in BGS surveys and number of galaxies in our sample. 

The sample volume obtained using this approach is:
\begin{equation}
V_m=\Omega{N_{FCRAO} \over N_{BGS}}{D^3 \over 3}=1.16\times D^3
\label{eq:Vm}
\end{equation}
\noindent
where D is maximum distance at which galaxy of given luminosity 
could be detected for a given flux limit, and $\Omega$ 
is the total solid angle covered by both BGS1 \citep{soi89} and BGS2 
\citep{san95} in {\it sterradian}.
$N_{FCRAO}/N_{BGS}$ is the ratio of the number of galaxies in our sample and
the number of galaxies in both BGS samples. 
BGS1 and BGS2 together contain 601 galaxies and cover 83\% of the sky.

\subsection{60 $\mu$m Luminosity Function \label{sec:60LF}}

The 60 $\mu$m luminosity represents the contribution from the IRAS 
60 $\mu$m band to the FIR luminosity \citep{helou88} and was 
calculated using the following relation from \citet{yun01} :
\begin{equation}
log L_{60\mu m}(L_{\odot})=6.014+2logD+logS_{60\mu m}
\label{eq:L60}
\end{equation}
\noindent
where D is the distance in Mpc and $S_{60\mu m}$ is the IRAS 60 $\mu$m
band flux density in Jy.

\begin{figure}[ht]
\plotone{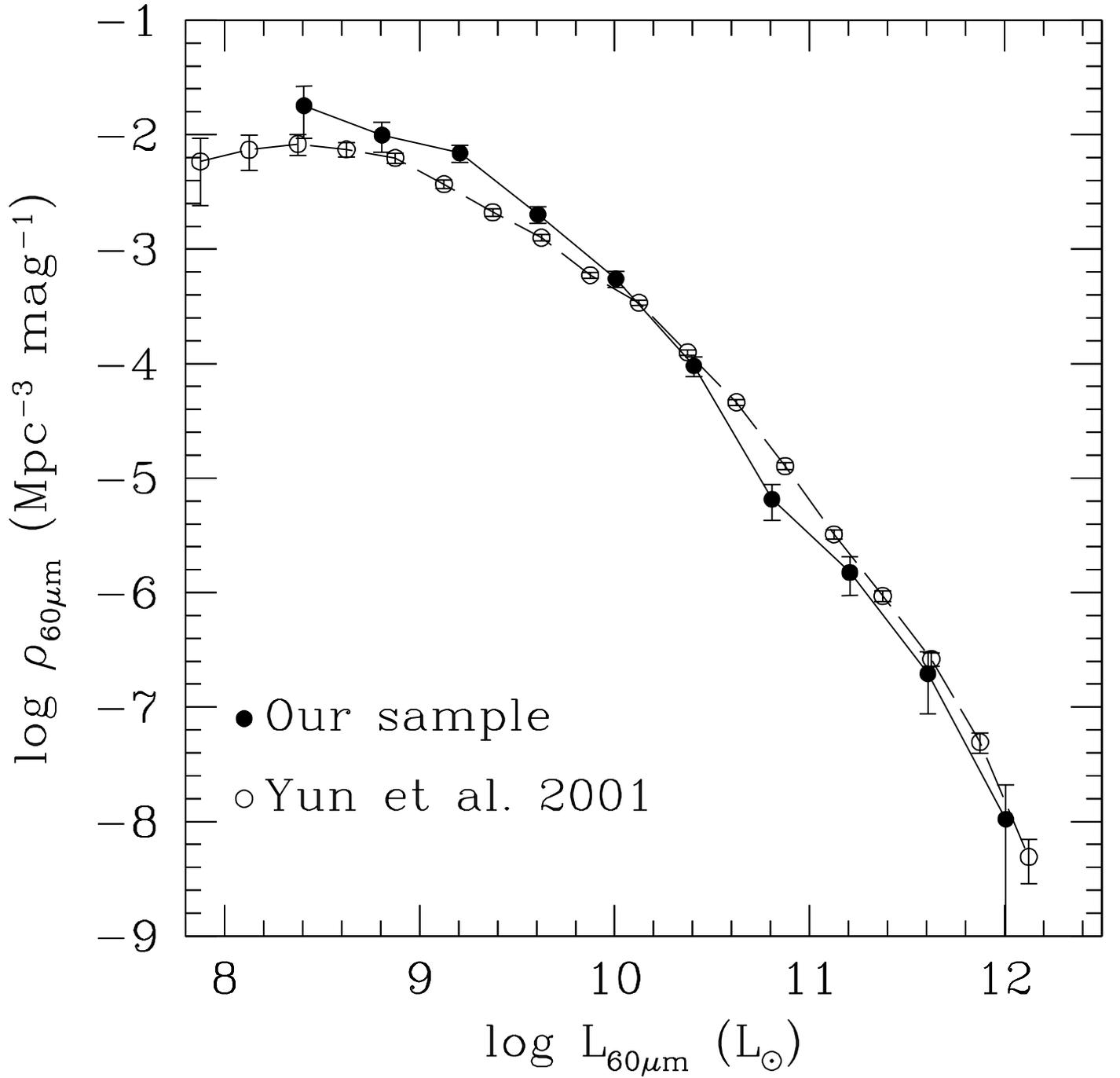}
\caption{
IRAS 60 $\mu$m LF obtained using our CO sample of 200 galaxies corrected for
the sample selection.  The comparison with the 60 $\mu$m LF derived from 
a much larger sample obtained by \citet{yun01} shows a good agreement.
\label{fig:60LF}
}
\end{figure}

As a way to verify our CO sample correction and to evaluate the effects of
various sample biases discussed in \S~\ref{sec:sample}, we first derive the 
60 $\mu$m LF for our CO-selected galaxies in order to 
compare it to the previously published 60 $\mu$m LF by \citet{yun01}. 
As shown on Figure~\ref{fig:60LF}, the agreement between 
the two 60 $\mu$m luminosity functions is excellent.  Yun et al. 
sample was selected with 2.5 times smaller flux density cutoff than 
the BGS and is thus 9 times larger in size.  While this is not a
complete verification of our derivation method, this agreement 
lends support to our assumption that our sample is a fair subset of 
the complete 60$\mu m$ selected sample.
The normalization factor could in principle be dependent on the 
60$\mu m$ luminosity.  The comparison of the 60$\mu m$ LFs suggests 
that this dependence is weak if any. 

About 13\% of our sample are Virgo cluster galaxies.  Again, the
60 $\mu$m luminosity function was constructed including and excluding 
Virgo galaxies. The resulting luminosity functions are very similar
-- see Figure~\ref{fig:COIRLF}. 
The similarity would arise if the cluster galaxies are distributed 
in luminosity bins in the same manner as the field galaxies and if the
Virgo galaxies contribute only a small fraction to each luminosity
bin.  Previous studies of Virgo galaxies have suggested that their
molecular gas contents are affected little by their cluster 
environment \citep[e.g.][]{kyn89}. 
Since we are interested in the total cold gas density of the local
volume regardless of the galaxy environment, we include all
Virgo galaxies in our analysis here on.

\subsection{CO Luminosity Function \label{sec:COIRLF}}

Encouraged by the success of deriving the 60 $\mu$m LF above for
the 200 galaxies in our sample, the
CO luminosity function is derived using the $^{12}$CO(1--0) flux 
with the same sample and volume correction. 
The CO luminosity is calculated as $L_{CO}=4 \pi D^2 
S_{CO}$, where D is the distance to the galaxy in Mpc and
$S_{CO}$\footnote{We adopt the definition $S_{CO}\equiv \int 
I_{CO}d\Omega$, following the same notation as \citet{yng95}.} 
is the velocity integrated CO flux in Jy km s$^{-1}$. 
The computed CO luminosity for our sample galaxies ranges between 
$10^4$ and $10^8$ Jy km s$^{-1}$ Mpc$^2$.  They are divided into 
10 bins of unit magnitude (0.4 wide in a log scale).  There are larger
uncertainties at the low luminosity end because only a small number
of sources occupy these bins. 

\begin{figure}[ht]
\plotone{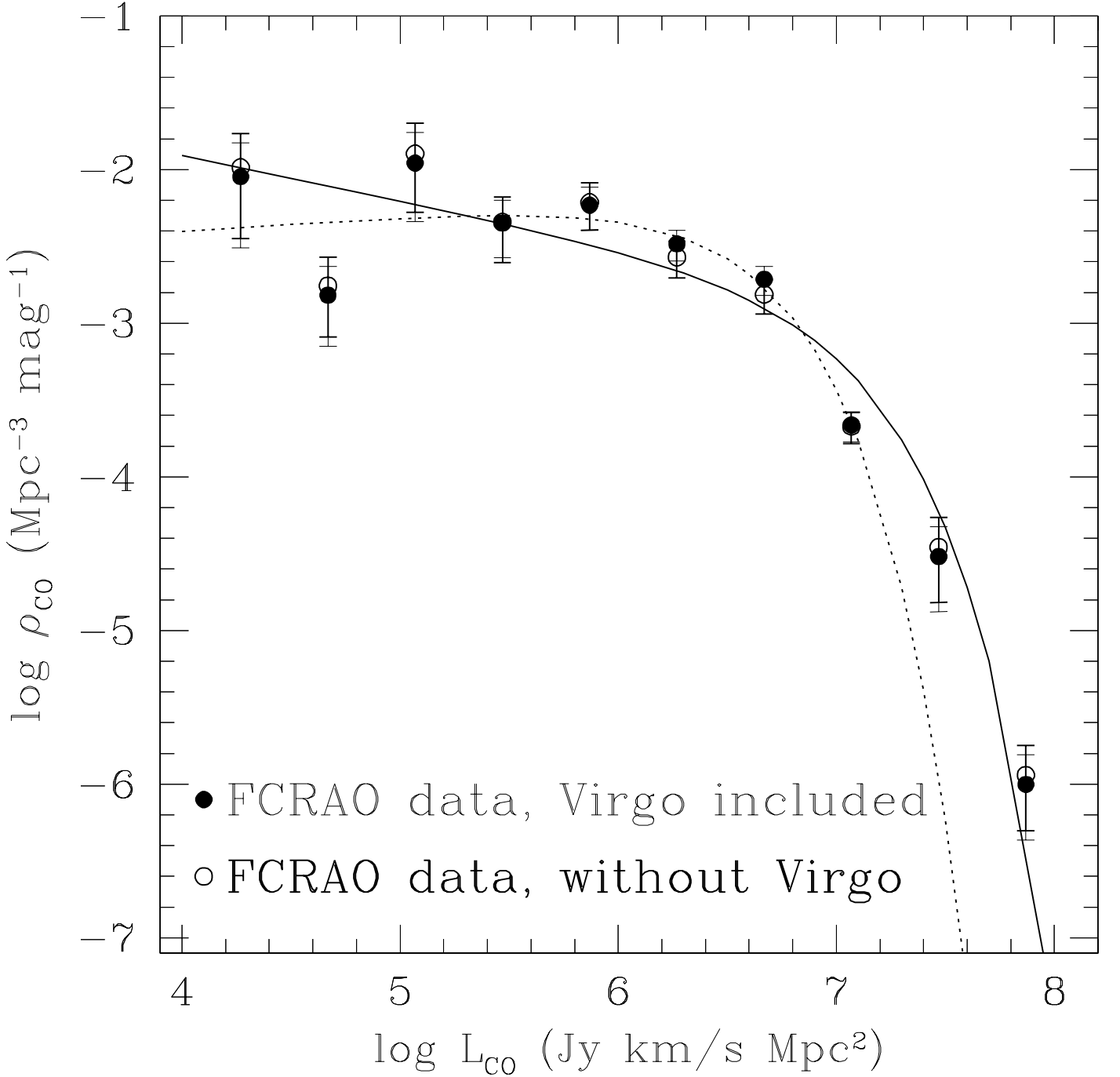}
\caption{
The CO luminosity function, derived using the same sample and
volume correction 
as the IRAS 60 $\mu$m LF shown in Figure~\ref{fig:60LF}.  
The solid line is the best fit Schechter function derived using all 10 bins.
The CO LF follows a Schechter function shape closely
with a faint end power-law index $\alpha=-1.30$ and the characteristic
luminosity $L^*=9.8 \times 10^6$ Jy km s$^{-1}$ Mpc$^{2}$. 
The dotted line shows the best fit Schechter function 
derived excluding the two highest
luminosity bins (see \S~\ref{sec:COIRLF}). The power-law index for this fit is
$\alpha=-0.90$ and $L^*=3.3\times 10^6$ Jy km s$^{-1}$ Mpc$^{2}$.
\label{fig:COIRLF}
}
\end{figure}

The resulting CO luminosity function is shown in Figure~\ref{fig:COIRLF}. 
It is well described by a Schechter function which has an exponential 
cutoff at high luminosity end.  The low luminosity end power-law slope
is quite flat, similar to the 60 $\mu$m LF (see \S~\ref{sec:discussion}). 
To quantify these properties, the characteristic 
parameters are derived for the Schechter function \citep{sch76} of the form
\begin{equation}
\phi (L)d(L)=\phi^*
(\frac{L}{L^*})^{\alpha}exp(-\frac{L}{L^*})d(\frac{L}{L^*})
\label {eq:Schechter}
\end{equation}
\noindent
where $\phi^*$ is the normalization factor, $\alpha$ is the low
luminosity end power-law slope, and $L^*$ is the characteristic 
luminosity at which the exponential cutoff develops.
Since we are using logarithmic intervals to construct LF, we need to change
this equation to an appropriate form:
\begin{equation}
\rho (L)=\rho^*(\frac{L}{L^*})^{\alpha+1}exp(-\frac{L}{L^*})\ln{10}
\label {eq:LSchechter}
\end{equation}
\noindent
in this case $\rho(L)=\phi(L) d(L)/d(\log(L))$, and it is calculated using
eq. (~\ref{eq:1/vm}).  
The derived best fit Schechter parameters are:
$\rho^*=(0.00072 \pm 0.00035)$ Mpc$^{-3}$ mag$^{-1}$,
$\alpha=(-1.30 \pm 0.16)$, and 
$L^*=(1.0 \pm 0.2)\times 10^7$ Jy km s$^{-1}$ Mpc$^2$
($M^*=9.4\pm 1.9\times 10^9 M_\odot$ using the standard
CO-to-H$_2$ conversion factor -- see \S~\ref{sec:gasdensity}).

The single Schechter function fit using all 10 luminosity bins
produces a rather poor fit with $\chi ^2=50$.  This poor fit 
is also obvious in Figure~\ref{fig:COIRLF} as the single Schechter 
function fit shown with a solid line is only marginally consistent
with the data points and their formal uncertainties.  It is
possible that a Schechter function is intrinsically a poor functional
form for the CO LF.  On the other hand, a Schechter function 
successfully describes many other LFs including the HI LF
\citep[e.g.][]{sch98}.  There is also a theoretical basis in 
that it is the functional form associated with the halo mass 
function in the Press-Schechter formulation of the structure 
formation \citep[][]{Press74}.

One possible explanation for such a large $\chi ^2$ value 
is that there are two (or more) distinct populations of galaxies 
in the sample, similar to the situation with the FIR luminosity
function as noted by \citet{yun01}.  In the FIR, the light-to-mass
ratio for the starburst population is systematically
enhanced by 1-2 orders of magnitudes over the field population.  
The CO luminosity
is also thought to be elevated significantly among the intense
nuclear starburst systems \citep{sco97,dow98}, and this may
materialize as two distinct populations in the CO LF as well. 
There are 14 mergers/starburst systems in our sample, and eight of 
these fall within the last two bins and account for about 40\% 
of the galaxies in these bins. 
When a best fit Schechter function is derived using only the first 
8 bins dominated by normal field galaxies, the $\chi ^2$ value drops 
to 13, which is a significant improvement. 
The Schechter function fit for the bins dominated by normal galaxies gives:
$\rho^*=(0.003 \pm 0.001)$ Mpc$^{-3}$ mag$^{-1}$,
$\alpha=(-0.90 \pm 0.15)$, and 
$L^*=(3.3 \pm 0.6)\times 10^6$ Jy km s$^{-1}$ Mpc$^2$
($M^*=3.1\pm 0.6\times 10^9 M_\odot$).

The difference between all Schechter parameters for the two
limiting situations of treating the non-detected galaxies 
as upper limits or zero detections is smaller than 2\%. Such a small
difference is expected since we have less than 4\% upper limits in our
sample. 
Similarly, including or excluding Virgo galaxies has only a minor
impact, as discussed above (see \S~\ref{sec:60LF}).

\subsection{CO LF from the $B$-band Selected Galaxies \label{sec:COBLF}}

As an independent check of our CO LF derived using the FIR selection, we
also derive a CO luminosity function using an optical $B$-band 
selected sample.  The method of deriving the CO LF is similar 
to the method used for the infrared selected galaxies: we use 
the $1/V_m$ method and rescaling of the number of
galaxies using a complete sample, which is the Revised Shapley-Ames 
(RSA) catalog by \citet{sat81} in this case.  A declination 
cut off of $\delta > -25^\circ$ and a Galactic longitude limit 
of $|b| > 10^\circ$ are adopted as part of the sample definition.
This makes the total surveyed area equal to 60\% of the sky.
Our brightness selection criterion is $B_T \le 12.0 ^{mag}$.

Accounting for different Hubble types correctly is a difficult issue.
Since CO emission is generally very weak or undetected among
early type galaxies, only the spiral and irregular galaxies are
included in the analysis.  
The examination of the differential number counts for our $B$-band
selected sample yields a power-law index $\alpha=-1.35$, suggesting
a reasonably good uniformity and completeness.  The mean $<V/V_m>$
ratio for the sample is 0.33 (see Fig.~\ref{fig:vvm}), similar
to the FIR selected sample. 
Although we adopt the RSA catalog as a complete sample, 
a few galaxies satisfying our selection criteria are 
missing in the RSA \citep[see][]{con87}.

\begin{figure}[ht]
\plotone{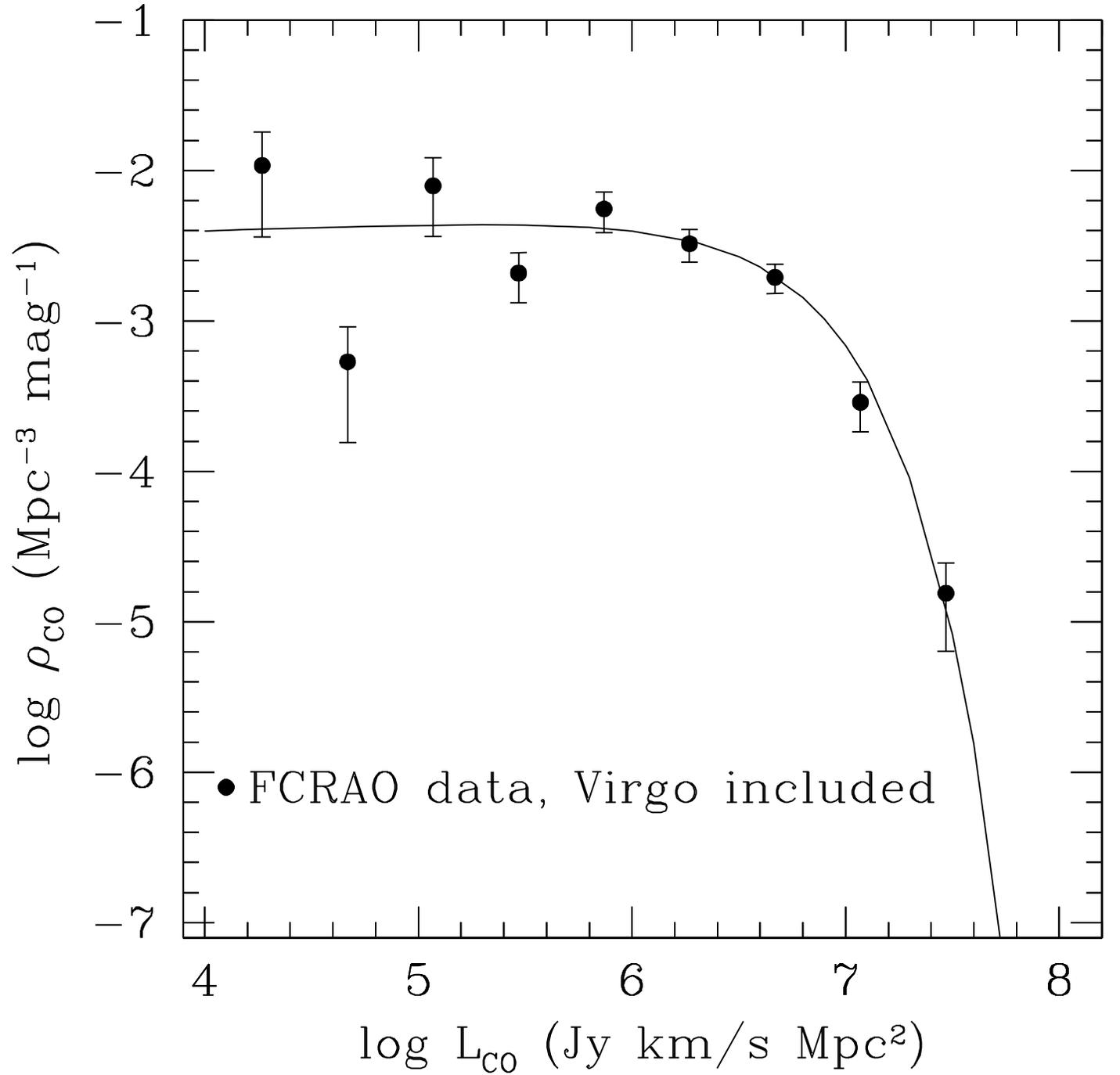}
\caption{
The CO luminosity function obtained for the optical 
$B$-band selected sample derived
from the RSA catalog as the parent sample.  The solid line shows
the best fit Schechter function (see text). 
\label{fig:COBLF}
}
\end{figure}

The $B$-band magnitudes are taken from the RSA while 
the CO fluxes are taken from the Survey.  There are 133 galaxies in 
the CO Survey that satisfy the selection criteria while the parent
RSA catalog galaxies satisfying the selection criteria are 252 in total.
In this case, CO fluxes for 13 galaxies (10\%) are upper limits.
The CO luminosity function derived for this $B$-band selected
sample is shown on the Figure~\ref{fig:COBLF}. 
This LF is very similar to CO LF from the FIR selected galaxies 
but with a lower characteristic luminosity $L^*$ and a
larger scatter on the low luminosity end. 
The lower characteristic luminosity $L^*$ is
a direct consequence of the absence of mergers in this sample.
The best fit Schechter parameters for the $B$-band derived CO luminosity
function are: $\rho^*=(0.0021 \pm 0.0009)$ Mpc$^{-3}$ mag$^{-1}$,
$\alpha=(-1.0 \pm 0.2)$, and 
$L^*=(4.8 \pm 1.1)\times 10^6$ Jy km s$^{-1}$ Mpc$^2$
($M^*=4.5\pm 1.0\times 10^9 M_\odot$).

\bigskip

\section{Discussion \label{sec:discussion}}

\subsection{Uncertainties and biases in the CO LF \label{sec:bias}}

\begin{figure}[ht]
\plottwo{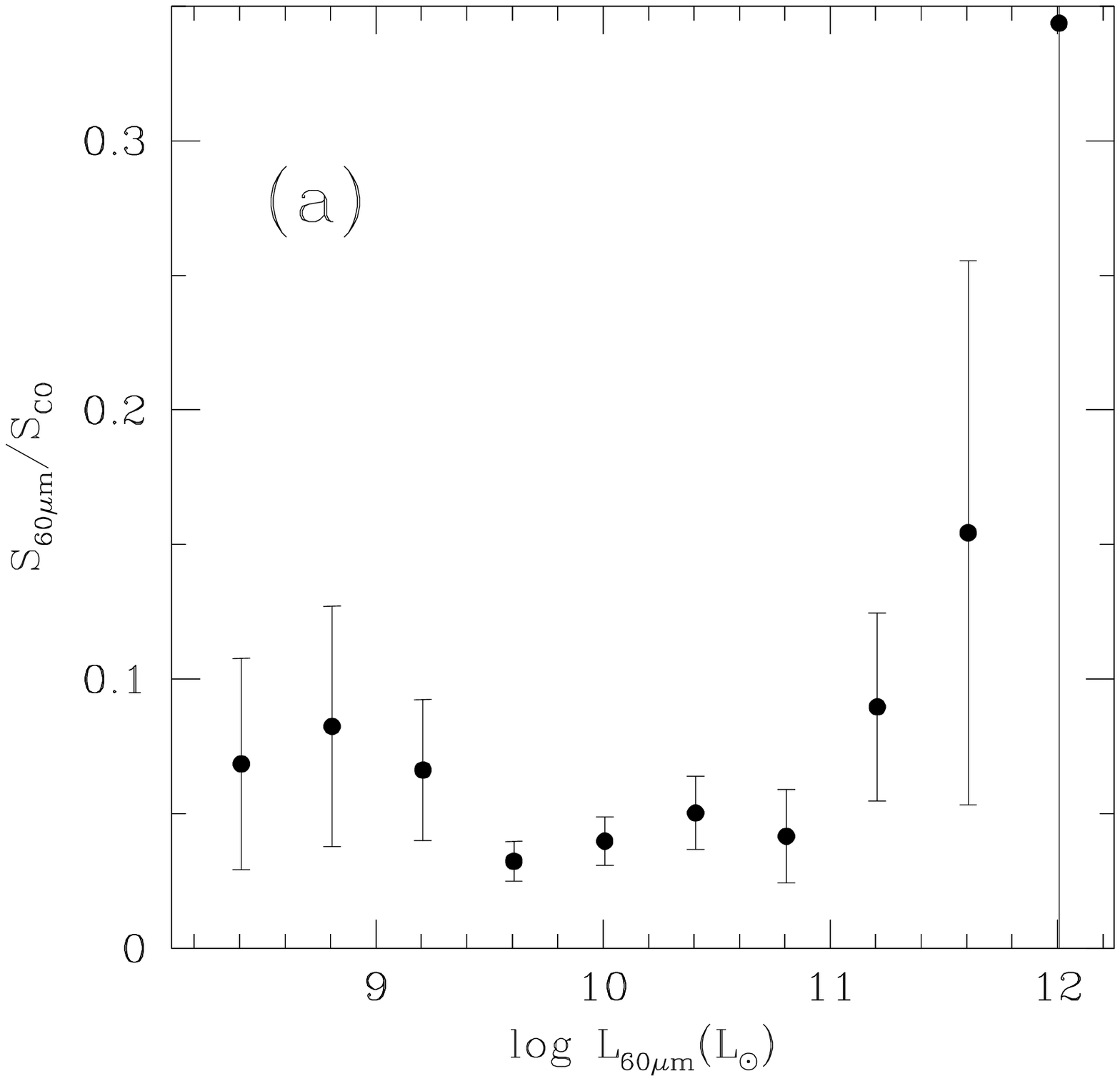}{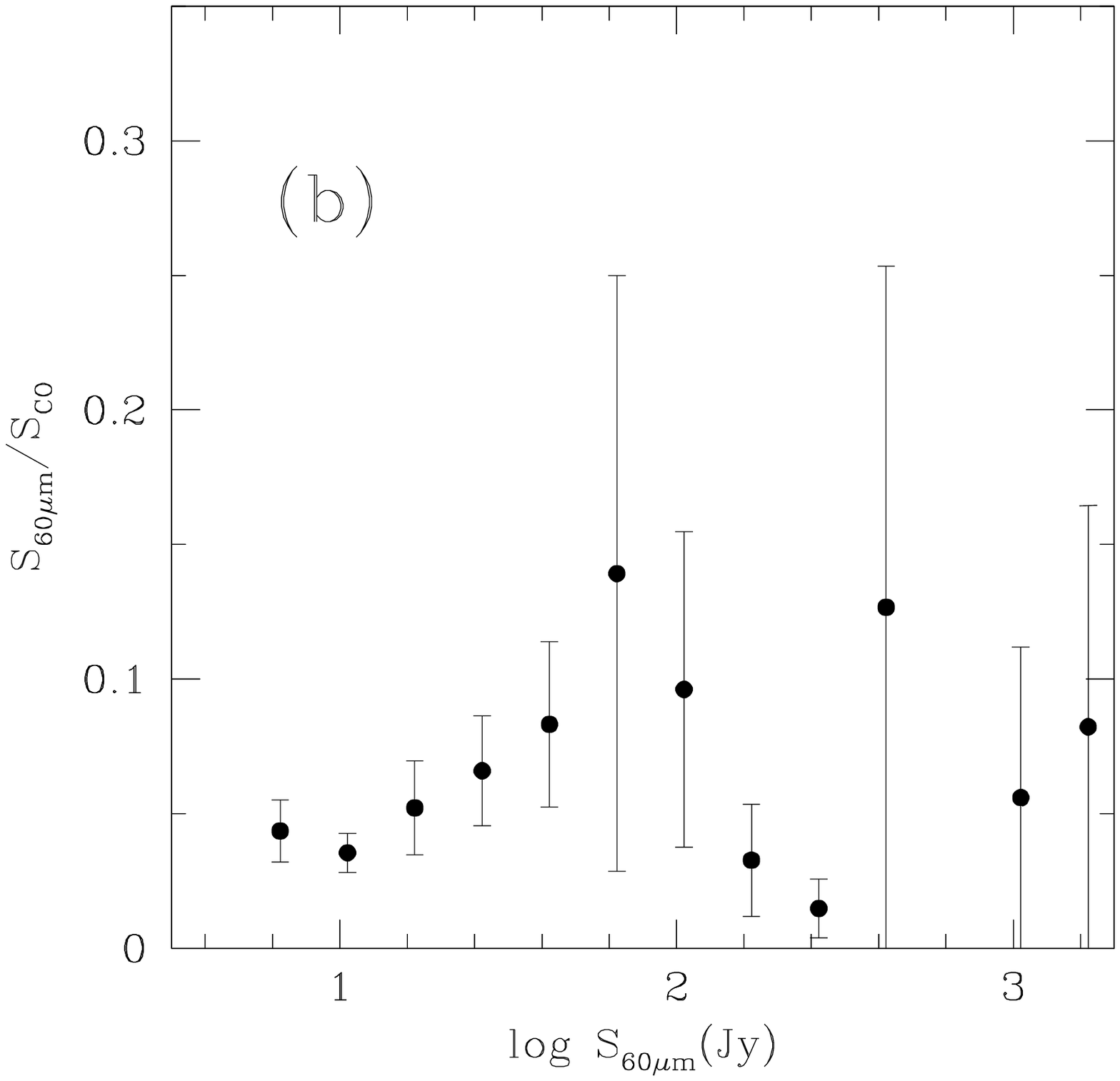}
\caption{
(a) Average $S_{60\mu m}/S_{CO}$ vs. $L_{60\mu m}$. 
High luminosity bins have higher values of $S_{60\mu m}/S_{CO}$, but
they also suffer from the small number statistics.
(b) Average $S_{60\mu m}/S_{CO}$ vs. $S_{60\mu m}$. The absence of 
any clear trend in either plots suggests that our IRAS selection
does not introduce any obvious bias to the CO measurements.
\label{fig:60/CO}
}
\end{figure}

One major source of uncertainty in the derived CO luminosity
function is that it is derived indirectly using the FIR and 
optical $B$-band selection functions.  A direct derivation from
the observed CO properties is possible in principle, but the  
complication associated with the {\it a priori} unknown
line widths adds a significant uncertainty.  Instead we take advantage
of the known tight correlation between FIR and CO luminosity
in deriving the CO LF for our sample.  The 1.4 GHz radio
luminosity function derived by \citet{yun01} using the radio-FIR
correlation and IRAS 60 $\mu$m flux density agrees very well
with the radio LFs derived directly \citep{con91,con02},
giving some assurance to this technique.  
We have examined whether any systematic trends (thus a bias) 
exist in the $S_{60\mu m}/S_{CO}$ ratio as a
function of the 60 $\mu$m flux density and luminosity.
As shown in Figure~\ref{fig:60/CO}, little trend is seen
in this ratio, and the known linear correlation seems to 
hold well within the uncertainties for the entire range of 
flux density and luminosity.  

Deriving a luminosity
function for one wavelength using a selection function at
another wavelength may be robust enough to work even if the
two quantities are correlated in a non-linear way.  
There is a known correlation between FIR, CO, and optical
$B$-band luminosity for late type galaxies, but the
correlations involving the $B$-band luminosity are shown to be
non-linear \citep[see][]{yng89,per97}.  Yet, the CO luminosity
function derived using the FIR selection function 
(Figure~\ref{fig:COIRLF}) is in excellent agreement with the
CO LF derived using the $B$-band selection function 
(Figure~\ref{fig:COBLF}).  The selection bias is present
as some of the most CO luminous galaxies (merger starbursts) are missed
in the $B$-band selected sample.  Nevertheless the agreement is
striking given the non-linear dependence between CO and $B$-band
luminosity.

Another source of a significant uncertainty is the uniformity of
sampling and correctly accounting for the selection function for
the sample of galaxies which is not complete by the adopted
selection criteria.  Both the differential source count
and the $<V/V_m>$ analysis suggest a non-uniform distribution
of the sample galaxies (see \S~\ref{sec:sample}), and a slight bias 
towards galaxies brighter at 60$\mu m$ may be responsible for this
effect.  The large scale structure also have influence in lowering 
$<V/V_m>$ value for our sample. The non-trivial nature of
these effects are demonstrated by the fact that including and excluding
Virgo cluster galaxies from the sample make little difference
to the differential source count and the $<V/V_m>$ values.

\subsection{Local Cold Gas Mass Density \label{sec:gasdensity}}

CO emission is a commonly used tracer of molecular hydrogen 
because CO is one of the most abundant  
molecules in cold ISM and because its excitation 
in astrophysical conditions is determined by collisions with
hydrogen molecules.  Using the CO LF derived above, we can for 
the first time estimate the total mass density of molecular
gas in the local volume.
Adopting $N(H_2)/I(CO)=3 \times 10^{20}$ cm$^{-2}$ 
[K km s$^{-1}$]$^{-1}$ \citep[see review by][]{ysc91},  
\begin{equation}
M(H_2)=1.18\times 10^4 S_{CO} D^2 M_{\odot}
\label{eq:mh2}
\end{equation}
\noindent 
where $S_{CO}$ is total CO flux of a galaxy 
in Jy km s$^{-1}$ and $D$ is luminosity distance in Mpc.

\begin{figure}[ht]
\plotone{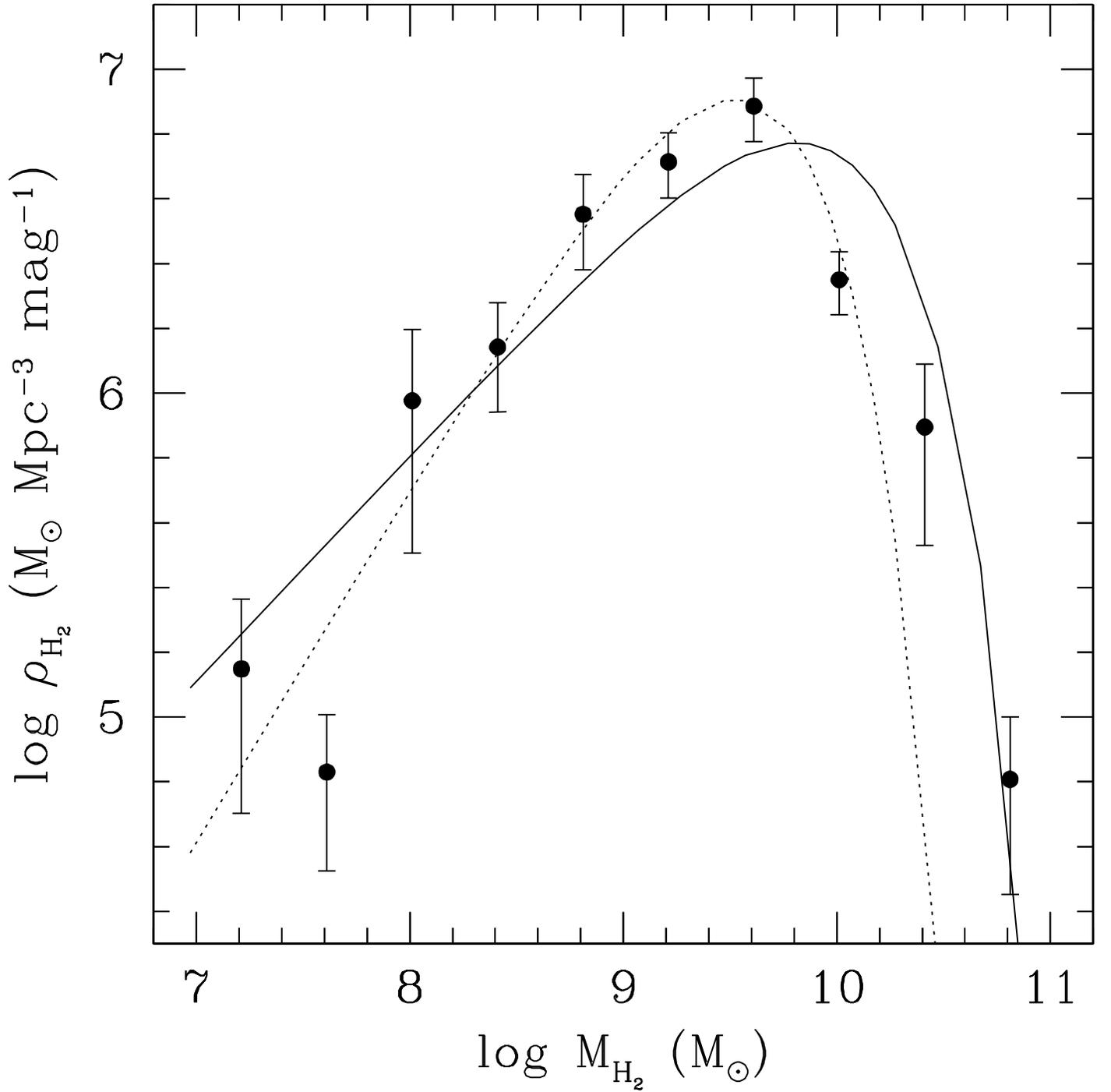}
\caption{
Molecular mass density contribution per magnitude of $H_2$ mass.
The H$_2$ mass bins near the $M^*$ value, log $M_{H_2}\sim 9.5$,
dominate the contribution to the total mass density.
The solid and dotted lines correspond to the same lines shown
in Figure~\ref{fig:COIRLF}.
\label{fig:h2mass}
}
\end{figure}

The molecular mass density in the local volume contributed by each
luminosity bin is 
shown in Figure~\ref{fig:h2mass}.  The dominant contribution to 
the mass density comes from galaxies around $L^*$ as expected.
The summation over the 10 bins gives 
$\rho_{H_2}=\sum M_{H_2}/V_m=(2.4 \pm 0.7) \times 10^7 M_{\odot}$ Mpc$^{-3}$. 
Integration of the LF using the Schechter parameters 
obtained in previous section gives a little bit smaller values:
$\rho_{H_2}=(2.2 \pm 1.1) \times 10^7 M_{\odot}$ Mpc$^{-3}$ for the fit trough
all 10 bins, and $\rho_{H_2}=(2.2 \pm 0.9) \times 10^7 M_{\odot}$ Mpc$^{-3}$
for fit trough first 8 bins. The second fit is more realistic, since it fits
much better bins that dominate contribution to the total mass, i.e. bins 
around $L^*$.
The uncertainty stated is $\pm 1\sigma$. The  systematic uncertainty, 
which includes uncertainties in the flux measurements, distance 
determinations, and CO-to-H$_2$ conversion, is probably larger.
Unless otherwise is stated, we adopt $\rho_{H_2}=(2.3 \pm 0.9) \times
10^7 M_{\odot}$ Mpc$^{-3}$ as an average value between values 
obtained from the fit and the direct summation. 
Since the dependence of the gas mass density on the Hubble
constant ($h\equiv H_0/{100}$ [km
s$^{-1}$ Mpc$^{-1}$]$^{-1}$) is linear, this result can be written as
$\rho_{H_2}=(3.1 \pm 1.2)\times 10^7 h M_{\odot}$ Mpc$^{-3}$.

The molecular hydrogen mass density in the local volume obtained from 
the $B$-band selected sample is $\rho_{H_2}=(3.1\pm 0.9) \times 10^7h 
M_{\odot}$ Mpc$^{-3}$ using a direct summation of the 
contribution of each galaxy and $(3.1 \pm 1.5) \times 10^7h M_{\odot}$
Mpc$^{-3}$ using the best fit Schechter parameters. 
Values for the local molecular mass density obtained
from the FIR selected sample and the $B$-band selected sample are in
good agreement.  Since we have better statistics for the 
larger FIR selected sample, we adopt the local molecular gas mass 
density derived from the FIR selected sample here on.
Using observed H$_2$/HI mass ratio for different
morphological types of galaxies and fraction of each morphological type
\citet{fuk98} estimated a similar value for $\rho_{H2}$.   

 In comparison, \citet{zwa97} estimated the local atomic gas mass density
of $\rho_{HI}=(5.8 \pm 1.2)\times 10^7 h M_{\odot}$ Mpc$^{-3}$ 
from their Arecibo HI strip survey while \citet{rab93} derived
$\rho_{HI}=(4.8 \pm 1.1) \times 10^7 h M_{\odot}$ Mpc$^{-3}$ 
from a sample of optically selected galaxies.  Therefore, the 
molecular gas mass density in the local volume
is about 50-65\% of the atomic mass density, and  molecular gas  
represents a significant component of the total mass density of the
neutral gas.  
The HI masses for 176 galaxies in our sample are 
known \citep{yng95}, and we have computed the HI mass density in 
the local volume using the same 
procedure as for the $H_2$ mass density. A summation 
over the HI mass weighted by $V_m$ gives a value 
consistent with the value obtain by \citet{zwa97}. 

\begin{figure}[ht]
\plottwo{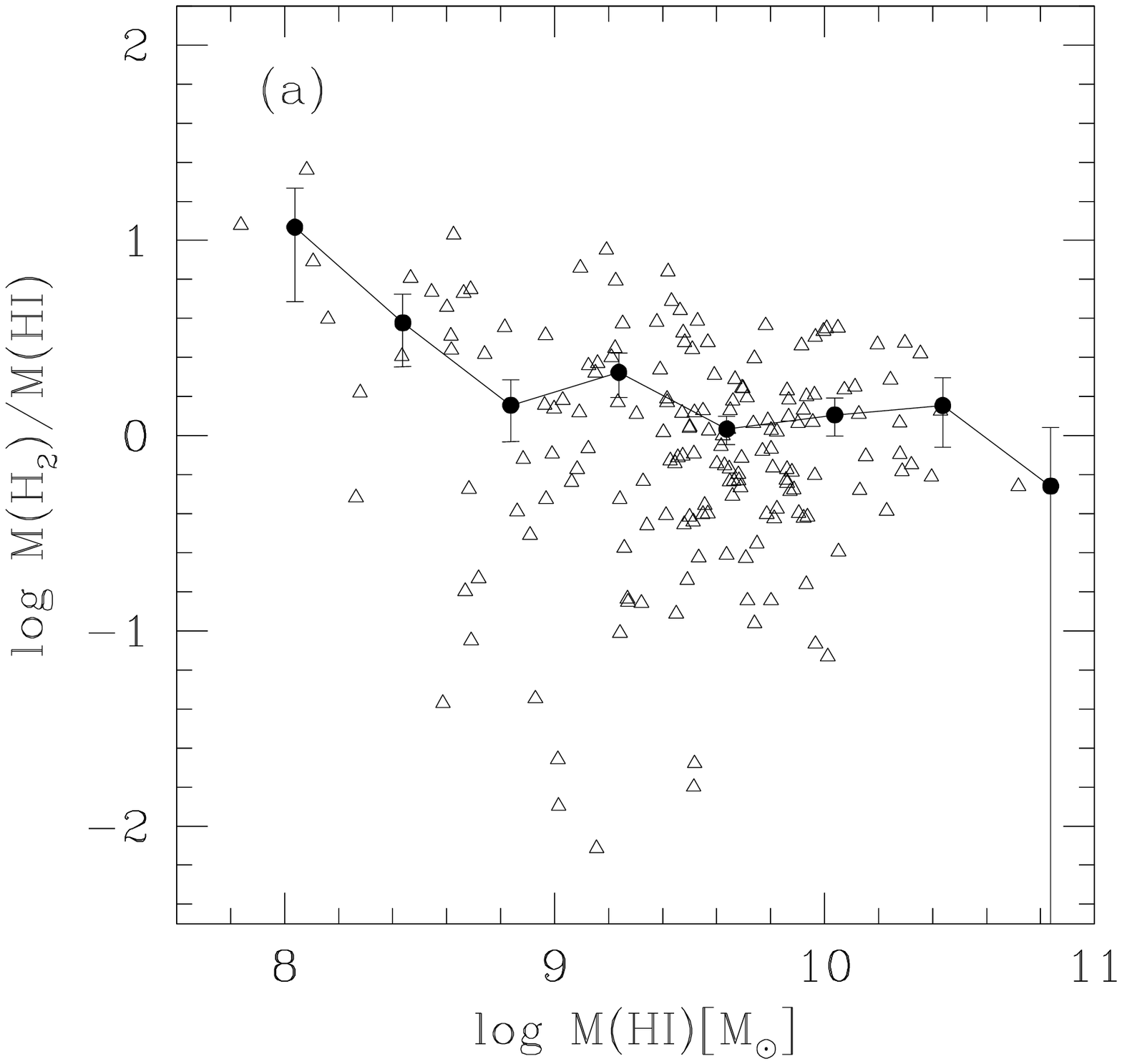}{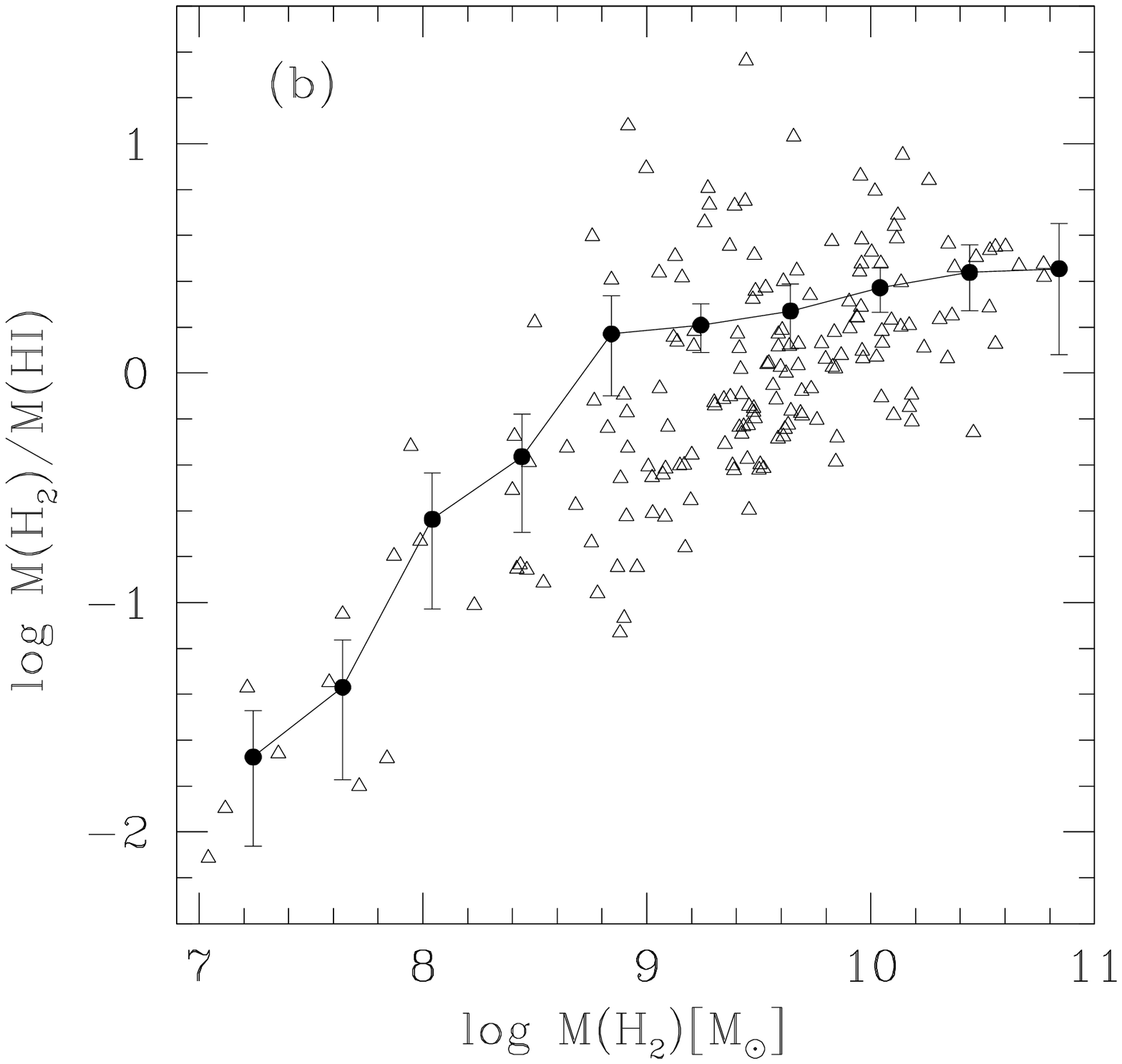}
\caption{
The molecular to atomic hydrogen mass ratio are shown as a function
of (a) $M_{HI}$ and (b) $M_{H_2}$.  Individual galaxies are shown
as triangles while the bin averages are shown with filled circles
and connected in solid lines to accentuate the
trend of growing $M(H_2)/M(HI)$ ratio for larger $H_2$ masses.
\label{fig:ratios}
}
\end{figure}

We examined our sample for any trends in $M_{H_2}/M_{HI}$ 
ratio vs. $M_{HI}$ and $M_{H_2}$. For the range of HI masses,
$8.8 < log~M_{HI}(M_\odot) < 11$, the ratio of molecular to neutral
atomic hydrogen masses is on average around unity with 
considerable scatter (see Fig.~\ref{fig:ratios}a).  Masked in
the large scatter is a possible trend
in $M_{H_2}/M_{HI}$ as a function of
Hubble type \citep[see][]{yng89}.  A clearer trend of increased 
$M_{H_2}/M_{HI}$ ratio with increasing H$_2$ mass is seen in
Figure~\ref{fig:ratios}b.  For $log~M_{H_2}(M_\odot) < 8.5$,
the average of this ratio is below 0.5 while the ratio jumps to
around 2 near $M_{H_2}\sim 10^9 M_\odot$.  Above H$_2$ mass of
$10^9 M_\odot$, on average galaxies have more molecular than atomic gas.

Using the molecular gas mass density derived here and the value of HI mass 
density from \citet{zwa97} we derive a total cold gas mass density 
at present epoch as a fraction of the critical density 
$\Omega_{HI+H_2}=(3.2 \pm 0.8) \times 10^{-4}h^{-1}$. 
Supposing that He contributes 25\% of the total gas mass density, 
we add 33\% to the derived value of $\Omega_H$, which gives 
$\Omega_{HI+H_2+He}\equiv \Omega_{gas}=(4.3 \pm 1.1) \times 10^{-4}h^{-1}$. 
These values for total neutral gas mass density are 15\% smaller if we use 
\citet{rab93} value for the HI mass density. 

The present baryon density estimated from the D/H ratio and Big Bang
nucleosynthesis is $\Omega_b = 0.02 \pm 0.002 h^{-2}$ \citep{bur01} while 
the estimate from the cosmic microwave background anisotropy is also 
around $\Omega_b\simeq 0.02 h^{-2}$ \citep{deb02}.  The stellar mass
contents account for $\Omega_* \simeq 0.0025 h^{-1}$ 
(Cole et al. 2001; but also see Benson, Frenk, \& Sharples 2002).  
We conclude that the cold gas content of late type galaxies is 
around 20\% of the stellar mass content and about 2\% of the
total baryonic content in the current epoch.

\bigskip

\subsection{CO-to-H$_2$ conversion \label{sec:Xfactor}}

The lack of electric dipole moment makes direct observations of
molecular hydrogen difficult in general, and studying the 
spatial extent and molecular gas mass requires
another tracer.  Highly abundant, chemically robust, and
easily excited by collision with H$_2$ molecules, 
CO is the most commonly used tracer of molecular
gas.  The CO (1--0) transition is optically thick under most
astrophysically interesting conditions, making it relatively
insensitive to metallicity and abundance effects.  The two
key excitation parameters of density and temperature have
a nearly canceling effect, making CO a fairly reliable
tracer of H$_2$ in a broad range of physical conditions 
\citep[see reviews by][]{mal88,ysc91}.

The derived ratios of $N(H_2)/I(CO)$ range between 
$(1-5) \times 10^{20}$ cm$^{-2}$ [K km s$^{-1}$]$^{-1}$ 
\citep{blo86,dic86,scs87}.
We adopt a constant CO-to-H$_2$ conversion factor of 
$N(H2)/I(CO)=3\times 10^{20}$ cm$^{-2}$ [K km s$^{-1}$]$^{-1}$ 
\citep[see discussions  by][]{ysc91}.  \citet{dey90} show that
the H$_2$ mass estimates of galaxies from their CO 
luminosity are accurate to $\pm30\%$.  
\citet{ysc91} show that the CO-to-H$_2$ conversion for galaxies
of diverse morphology and metallicity are similar in
absolute value to the conversion in the Milky Way.
While the $H_2$ mass estimate
for an individual galaxy may be uncertain to about 30\%,
the $H_2$ mass estimate for an {\it ensemble} of
galaxies should be more reliable.  In most galaxies,
giant molecular clouds (GMCs) and cloud complexes dominate the total
molecular gas mass, and adopting a conversion factor consistent with
the values derived from the Galactic GMCs should yield the
most robust, mass-weighted estimates of gas masses.

Among the low metallicity galaxies such as Small Magellanic 
Cloud, CO abundance may become low enough to affect the standard
assumption of self-shielding and thermalization.  The derived
conversion factors are generally larger \citep[e.g.][]{wil95},
and low metallicity dwarf galaxies are often undetected entirely in CO.
Therefore the H$_2$ mass and mass density derived from the CO luminosity
are strictly lower limits since molecular gas from these galaxies
is missing from our analysis.  
On the other hand, the contribution by the low luminosity 
bins to our derived $H_2$ mass density is small (see 
Fig.~\ref{fig:h2mass}) as the majority of the contribution 
to the total mass density comes from bins around $L^*$.
Therefore, we can largely neglect the effect of 
low metallicity, low luminosity galaxies in the derivation of the total
molecular gas mass density for the local volume. 

As discussed already in \S~\ref{sec:COIRLF}, the CO emission may
be elevated for the FIR luminous galaxies with nuclear gas concentrations,
and the CO-to-H$_2$ conversion factor may be smaller than the
canonical value.  In the extreme environment of the nuclear
starburst regions, CO emission arises from a multi-phase medium
with sub-thermal excitation in the diffuse phase, and the
standard conversion factor may over-estimate the molecular gas
mass by a factor as large as 3 to 5 \citep{sco97,dow98}. 
Such merger/starburst galaxies contribute significantly only for
the top two luminosity bins in our CO LF, and these two bins contribute
less than 5\% to the the total molecular gas mass density. 
Therefore, the high CO luminosity conversion factor does not have a 
significant impact on the derived total molecular mass density.

Similarly, increased CO emission in the central 1 kpc of our Galaxy
and other galaxies has been suggested by recent observations
\citep[][S. H\"{u}ttemeister, private communication]{pag01}.
High angular resolution observations of gas-rich spiral galaxies
frequently reveal a distinct component in the central 1 kpc
\citep[see][]{saka99}, and some fraction of the 
total CO luminosity may arise from such a component.
On the other hand, among the Virgo spirals studied by \citet{kyn88},
a simple exponential distribution without a significant central
component offers a good fit for the
12 out of 14 galaxies whose CO emission is spatially well resolved
(i.e., CO detected at $\ge4$ positions along the major axis).
We made no attempt to account for enhanced CO emission in the 
central kpc as the required information is 
generally not available.  If the molecular gas properties of
Virgo spirals are typical of the late type field galaxies, then
the possible contribution by the enhanced nuclear CO emission in some
galaxies may not be substantial.

In summary, using the canonical CO-to-H$_2$ conversion relation
is problematic in some cases, such as low metallicity systems
or luminous nuclear starburst systems, and CO is a poor tracer of
molecular gas among low metallicity dwarfs.
For these reasons the local molecular gas mass density we infer is really
a lower limit.  However, since the majority of the mass contribution 
comes from $L^*$ galaxies whose molecular gas mass is dominated by
GMCs like our Galaxy, the use of the standard conversion factor
still offers a fairly reliable estimate of the total molecular gas density
for the local volume.

\bigskip

\section{Summary \label{sec:summary}}

Utilizing the largest available CO survey, the FCRAO Extragalactic CO Survey
\citep{yng95}, the CO luminosity function for the local volume is
derived using (1) a FIR selected sample of 200 galaxies
that satisfy $S_{60\mu m}> 5.24 Jy$ and (2) optical $B$-band selected
sample of 133 galaxies.  Although neither of the samples is 
complete in terms of the sample selection, a sampling function
is constructed using a well defined parent sample, and a non-parametric
CO luminosity function is derived from each sample.  By examining
the properties of the CO luminosity functions, we conclude:

\begin{enumerate}

\item The CO luminosity functions derived from the FIR and $B$-band
selected samples are reasonably well described by a Schechter
function.  The characteristic luminosity $L^*$ is around 
$L_{CO} \sim 10^7$ Jy km s$^{-1}$ Mpc$^2$. 
The low luminosity end of the CO luminosity function is 
($\alpha= -1.3~{\rm to}~-0.9$).  Similar values
are obtained for the two CO LFs derived using the two differently 
selected samples.  

\item The molecular gas mass density of the local volume is 
$\rho_{H_2}=(3.1\pm1.1) \times 10^7 h
M_\odot$ Mpc$^{-3}$ which is about 50-65\% of the HI gas mass density. 
This value is not strongly affected by variations in the CO-to-H$_2$
conversion factor for low and high luminosity galaxies since it is
dominated by the $L^*$ galaxies with $L_{CO} \sim  10^7$ 
Jy km s$^{-1}$ Mpc$^2$ ($M^*_{H_2}\sim 5\times 10^9 M_\odot$).
Its dependence on the global CO-to-H$_2$ conversion factor is linear
and should be secure to within a factor of 30\% or better.

\item Combined with the HI gas mass density,
we estimate the total cold gas mass density 
at the present epoch as a fraction of the critical density 
$\Omega_{HI+H_2}=(3.2 \pm 0.8) \times 10^{-4}h^{-1}$. 
When the He contribution is included, the cold gas mass density increase to
$\Omega_{HI+H_2+He}\equiv \Omega_{gas}=(4.3 \pm 1.1) \times 10^{-4}h^{-1}$. 
Therefore, the cold gas content of late type galaxies corresponds to
about 20\% of the stellar mass content and about 2\% of the
total baryonic content in the universe.

\end{enumerate}

\bigskip

\acknowledgements

The authors acknowledge insightful discussions with N. Katz, M. Heyer, S.
H\"{u}ttemeister, and others.  Comments by our referee, 
Martin Zwaan, were quite helpful for improving the manuscript.
This research has made use of the NASA/IPAC Extragalactic Database 
(NED) which is operated by the Jet Propulsion Laboratory, California Institute
of Technology, under contract with the National Aeronautics and 
Space Administration.  D. Keres gratefully acknowledge the partial research 
support provided by the Mary Dailey Irvine graduate research fellowship.
The Five College Radio Astronomy Observatory
is operated with the permission of the Metropolitan District Commission, 
Commonwealth of Massachusetts, and with the
support of the National Science Foundation under grant AST 97-25951.

\newpage

\newpage

\newpage

\newpage

\newpage

\newpage

\newpage

\newpage

\newpage

\begin{table}
\caption{Galaxies with $S_{60} > 5.24$ Jy not detected in the Survey
\label{tab:missing_galaxies}}

\hskip 2pt

\begin{tabular}{|c|c|c|c|c|c|c|c|}
\hline
Name& R.A.& dec & D & $S_{60}$& $S_{CO}$& Type& Ref.\\
& [h m s]& [d m s]& [Mpc]& [Jy]& [Jy km/s]& & \\
\hline
NGC 337& 00 59 50.3& -07 34 44& 23.7& 9.33& 140& Sd& u\\
\hline
NGC 470& 01 19 44.8& +03 24 35& 35.6& 7.09& 180& Sb& u\\
\hline
NGC 693& 01 50 31.0& +06 08 42& 21.9& 6.86& 70& S0/a& u\\
\hline
NGC 925& 02 27 17.0& +33 34 43& 9.5& 9.03& 566& Sd& 1\\
\hline
NGC 3227& 10 23 30.6& +19 51 54& 20.6& 8.32& 756& SAB (pec)& *\\
\hline
NGC 3486& 11 00 23.9& +28 58 30& 7.4& 6.24& 263& Sc& 1\\
\hline
NGC 4027& 11 59 28.4& -19 19 55& 19.5& 11.89& 60& Sdm& u\\
\hline
NGC 4526& 12 34 03.1& +07 41 59& 16.0& 5.63& 87& S0& 2\\
\hline
NGC 4532& 12 34 19.3& +06 28 07& 16.0& 8.93& 60& Im& u\\
\hline
NGC 4808& 12 55 48.9& +04 18 13& 16.0& 6.92& 100& Scd& u\\
\hline
NGC 6824& 19 43 40.9& +56 06 33& 49.0& 5.94& 30& Sb& u\\
\hline
Mrk 273& 13 44 47.5& +55 54 11& 153.8& 22.09& 62& Merg.& 3\\
\hline
\end{tabular}

\hskip 2pt

Coordinates are for J2000 and are taken from NASA Extragalactic Database. 
$S_{60}$ is IRAS 60 $\mu$m flux density taken from BGS1 and BGS2
except for NGC 925 where we used value from \citet{ric88} increased by 18\%.
References for $S_{CO}$ (column 8) are:
1.-\citet{sag93};
2.-\citet{saw89};
3.-\citet{san91};
u - upper limit from \citet{yng95}.
*-a new measurement of central 45$''$ obtained using the FCRAO 14-m
telescope in March 2002. NGC~3227 was not included 
in the Survey originally because this galaxy was confused with NGC~3226.

\end{table}

\end{document}